\documentclass[preprint]{emulateapj}
\usepackage{graphicx}
\usepackage{amsmath}
\usepackage[mathscr]{eucal}
\usepackage{gensymb}
\usepackage{subfigure}
\begin{document}
\title
{
Lyot-based Low Order Wavefront Sensor for Phase-mask Coronagraphs: Principle, Simulations and Laboratory Experiments
}
\author
{
Garima Singh$^{1,2}$, Frantz Martinache$^{3}$, Pierre Baudoz $^{2}$, Olivier Guyon$^{1}$, Taro Matsuo$^{4}$, Nemanja Jovanovic$^{1}$, Christophe Clergeon$^{1,2}$\\
$^1$\small Subaru Telescope, National Astronomical Observatory of Japan, Hilo, HI  96720, USA\\
$^2$\small Lesia, Observatoire de Paris-Meudon, 5 Place Jules Janssen, F-92195 Meudon Cedex, France\\
$^3$\small Observatoire de la C\^ote d'Azur, Boulevard de l'Observatoire, 06300 Nice, France\\
$^4$Kyoto University, Kitashirakawa-Oiwakecho, Sakyo-ku, Kyoto 606-8502 Japan\\
}
\email{singh@naoj.org} 
%
%
\begin{abstract}
High performance coronagraphic imaging of faint structures around bright stars at small angular separations requires fine control of tip, tilt and other low order aberrations. When such errors occur upstream of a coronagraph, they results in starlight leakage which reduces the dynamic range of the instrument. This issue has been previously addressed  for occulting coronagraphs by sensing the starlight before or at the coronagraphic focal plane. One such solution, the coronagraphic low order wave-front sensor (CLOWFS) uses a partially reflective focal plane mask to measure pointing errors for Lyot-type coronagraphs. 
 
To deal with pointing errors in low inner working angle phase mask coronagraphs which do not have a reflective focal plane mask, we have adapted the CLOWFS technique. This new concept relies on starlight diffracted by the focal plane phase mask being reflected by the Lyot stop towards a sensor which reliably measures low order aberrations such as tip and tilt. This reflective Lyot-based wavefront sensor is a linear reconstructor which provides high sensitivity tip-tilt error measurements with phase mask coronagraphs.

Simulations show that the measurement accuracy of pointing errors with realistic post adaptive optics residuals are $\approx~10^{-2}~\lambda/D$ per mode at $\lambda$ = 1.6~$\micron$ for a four quadrant phase mask. In addition, we demonstrate the open loop measurement pointing accuracy of $10^{-2}~\lambda$/D at 638~nm for a four quadrant phase mask in the laboratory. 
\end{abstract}
%
%
\section{Introduction}

The 40 milli-arcsecond diffraction limited resolution at $\lambda\simeq$~1.6~$\micron$ ($\lambda$=wavelength) provided by 8-meter class ground-based telescope is theoretically sufficient for the direct detection and characterization of extrasolar planets in the habitable zones of nearby stars ($<$ 30 pc). Their direct imaging, however, is affected by the rapidly changing atmosphere as well as optical imperfections and residual quasi static aberrations which limit the optical system's high contrast imaging capability. Accurate measurement and calibration of the wavefront is therefore required. 

With Adaptive Optics (AO), the ability of current telescopes to reach the diffraction-limit has been improved greatly and post processing techniques such as differential imaging have made it possible to identify faint companions at angular separation $\gtrsim$~10~$\lambda$/D from their parent star (D= Pupil diameter) (\citet{Marois2008} and \citet {Lagrange2009}). 

The newly developed high performance small inner working angle (IWA) coronagraphs employed on Extreme Adaptive Optic Systems (ExAO) are trying to image exoplanets within a few $\lambda$/D. At such small angular separations, the telescope pointing errors make it difficult to center the stellar light on the occulting mask which creates a halo of speckles around the occulter preventing detection of companions. In other words, these high performance coronagraphs are extremely sensitive to tip-tilt errors (\citet{Lloyd2005}; \citet{Shaklan2005}; \citet{Siva2005}; \citet{Belikov2006}; \citet{Guyon2006}). 

The absence of accurate pointing control degrades the coronagraph starlight rejection capability by allowing the planet's photons to mix with the starlight leaking around the coronagraph focal plane mask. Thus exoplanet direct imaging is often limited by how well low order wavefront aberrations upstream of a coronagraph are controlled and calibrated. 

To address this issue, the LYOT project \citep{Digby2006} proposed to re-image the starlight at the coronagraphic focal plane in order to track pointing errors. Current ExAO systems such as the Gemini Planet Imager (GPI) and the Subaru Coronagraphic Extreme Adaptive Optics (SCExAO) system use an adaption of the LYOT project's concept as outlined below. 

The Spectro-Polarimetric High-contrast Exoplanet Research (SPHERE) instrument for the Very Large Telescope (VLT) uses a Differential Tip-Tilt Sensor (DTTS) unit which is located as close to the coronagraphic focal plane mask as possible to minimize differential movement between elements. It measures the center of gravity to estimate the exact position of the beam on the coronagraph. Laboratory results of the DTTS have shown a centroid measurement precision of 0.14 mas/hour/axis if room temperature variations are kept below 0.5$\degree$C \citep{Baudoz2010}.

In contrast, GPI uses the starlight that passes through the focal plane mask and re-images the pupil on a Shack-Hartmann wavefront sensor in its calibration unit. GPI's laboratory result have shown a pointing accuracy of 2 mas in 20 seconds for an $8^{th}$ magnitude star for an apodized-pupil coronagraph \citep{Wallace2010}. 

The SCExAO system at Subaru Telescope is using the CLOWFS concept as described in \citet{Guyon2009} and \citet{Vogt2011}. The CLOWFS uses a dual-zone focal plane mask which blocks the center of the point spread function (PSF), partially reflects the wings of the PSF with a reflective annulus and allows off-axis sources to be transmitted towards the science camera. The recent laboratory demonstration of similarly designed CLOWFS on NASA's High Contrast Imaging Testbed at the Jet Propulsion Laboratory has shown the stabilization of tip-tilt with 0.001~$\lambda$/D residuals in closed loop at $\lambda$ = 808 nm with a phase-induced amplitude apodization (PIAA) coronagraph \citep{Kern2013}.

Despite the high level of tip/tilt control demonstrated on occulting coronagraphs, there is no solution yet in regards to tip/tilt sensing for phase mask coronagraphs (PMC). With the aim of imaging high contrast structures at 1~$\lambda$/D with phase mask-based coronagraphs, we introduce a new generation of CLOWFS which is compatible with PMC. It is based on a reflective Lyot stop (RLS) which re-images the diffractive focal plane mask. The new system is called the Lyot-based low order wavefront sensor (LLOWFS). 

A common property of all PMCs is that they redistribute the energy spatially in the pupil plane downstream of the coronagraph, canceling on-axis light and diffracting it outside the geometric pupil which is then blocked by a conventional Lyot stop. To control pointing errors with a PMC, we have modified the Lyot design, so that the Lyot stop reflects the unused starlight towards a low-order sensor that reliably measures tip-tilt errors. 

The RLS concept is presented in detail in section 2. The pointing error estimation is explained in Section 2.2. We describe a typical generic LLOWFS optical configuration, numerical simulation and the results in section 3. We then use realistic AO residuals as a simulation input to test the ability of our technique to measure the low order modes in section \ref{s:sim1}. Section \ref {s:lab} discusses the results of the preliminary implementation of the RLS with a four quadrant phase mask (FQPM) \citep{Rouan2000} in a coronagraphic testbed at LESIA, France. 
%
%
%
\section{Principle}
\subsection{A reflective Lyot stop wavefront sensor}
In its simplest form, a coronagraph is an occulting disk in the focal plane of the telescope blocking the central airy disc of the star, combined with a Lyot stop which reduces the stellar glare by eliminating the rest of the diffracting light allowing detection of companions or disk structures.

\begin{figure}[ht]
   \centerline{
        \resizebox{0.48\textwidth}{!}{\includegraphics{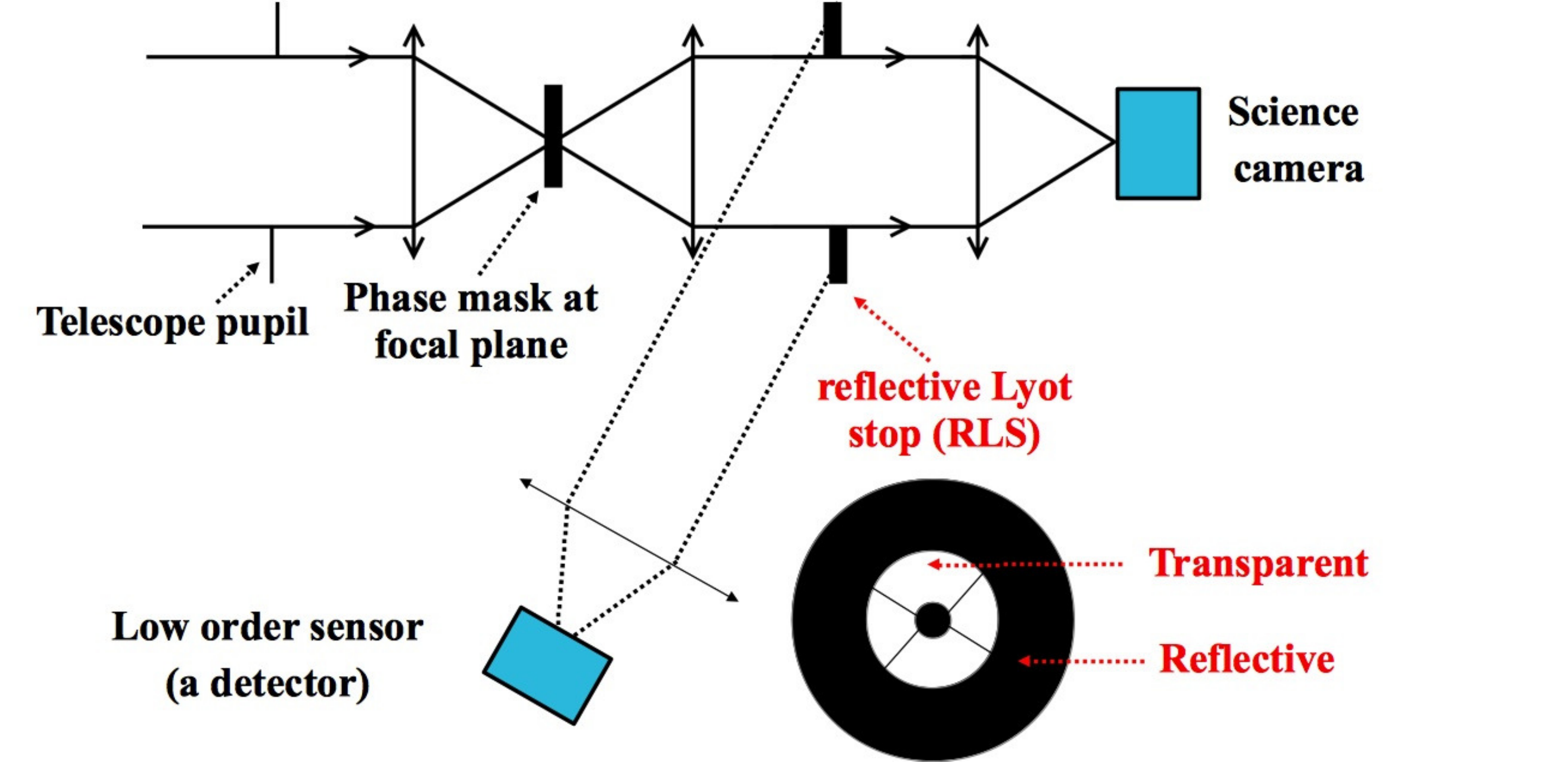}}}
   \caption{Schematic of the Lyot-based low order wavefront sensor (LLOWFS) consisting of a high performance coronagraph combined with a reflective Lyot stop (RLS) with a geometry that can be adopted according to the telescope pupil shape (central obstruction and spider arms) and diffraction pattern created by phase mask at focal plane.}
  \label{f:plotone}
\end{figure}

The technique of using a reflective focal plane mask as a means of measuring and correcting the low order aberrations such as the CLOWFS concept has enabled high contrast imaging in 1 - 2~$\lambda$/D region. However, for high performance PMCs, a reflective focal plane mask as used in CLOWFS is not applicable, hence a new solution is required. The LLOWFS instead accommodates the diffractive nature of the PMC focal plane mask in a manner illustrated in Figure~\ref{f:plotone}. At the focal plane of the telescope, a phase mask diffracts starlight in a re-imaged pupil plane. The starlight is then directed by the RLS towards a detector which is used for low-order sensing.

At this point it is possible to use the pupil plane image to drive the low order sensor as phase aberrations present in the intermittent focal plane are converted to intensity modulations via the phase mask used (e.g. the FQPM \citep{Rouan2000}, the Vortex \citep{Mawet2010}). However to maximize the signal-to-noise ratio and hence be able to operate on faint host stars, we conduct the wavefront sensing in a re-imaged focal plane which offers the highest photon density per pixel.

For the accurate detection of the focus aberration, a deliberate defocus is introduced in the position of the detector in order to eliminate the sign ambiguity with respect to which side of the sensor the focus aberration is. The defocus in the detector position ensures that the image plane is on a single side of the focus at all times. A detailed study of the retrieval process with a defocused image is presented in \citet{Guyon2009}.

As we operate with a slight defocus the flux density per pixel is reduced which only becomes an issue in the readout noise limited case. In addition we do not expect crosstalk to be introduced between the modes due to defocus but we have not quantified/investigated this thus far. 

\citet{Guyon2009} also demonstrate a detailed closed loop operation on focus and the astigmatisms and, in principle, LLOWFS should be able to measure other modes such as focus and the astigmatisms apart from just pointing errors. However, our key motivation is to address the pointing errors only for this body of work and the measurement of other low-order modes will be addressed in future work. 

The configuration of LLOWFS that we will discuss in this paper is based on the revised SCExAO testbed \citep{Jovanovic2013} which has been recently equipped with phase masks coronagraphs such as the Vortex, the eight octant phase mask \citep{Murakami2008a}, the phase-induced amplitude apodization with a variable focal plane mask \citep{Newman2013} and the four quadrant phase mask. 
 \begin{figure*}
   \centerline{
        \resizebox{0.8\textwidth}{!}{\includegraphics{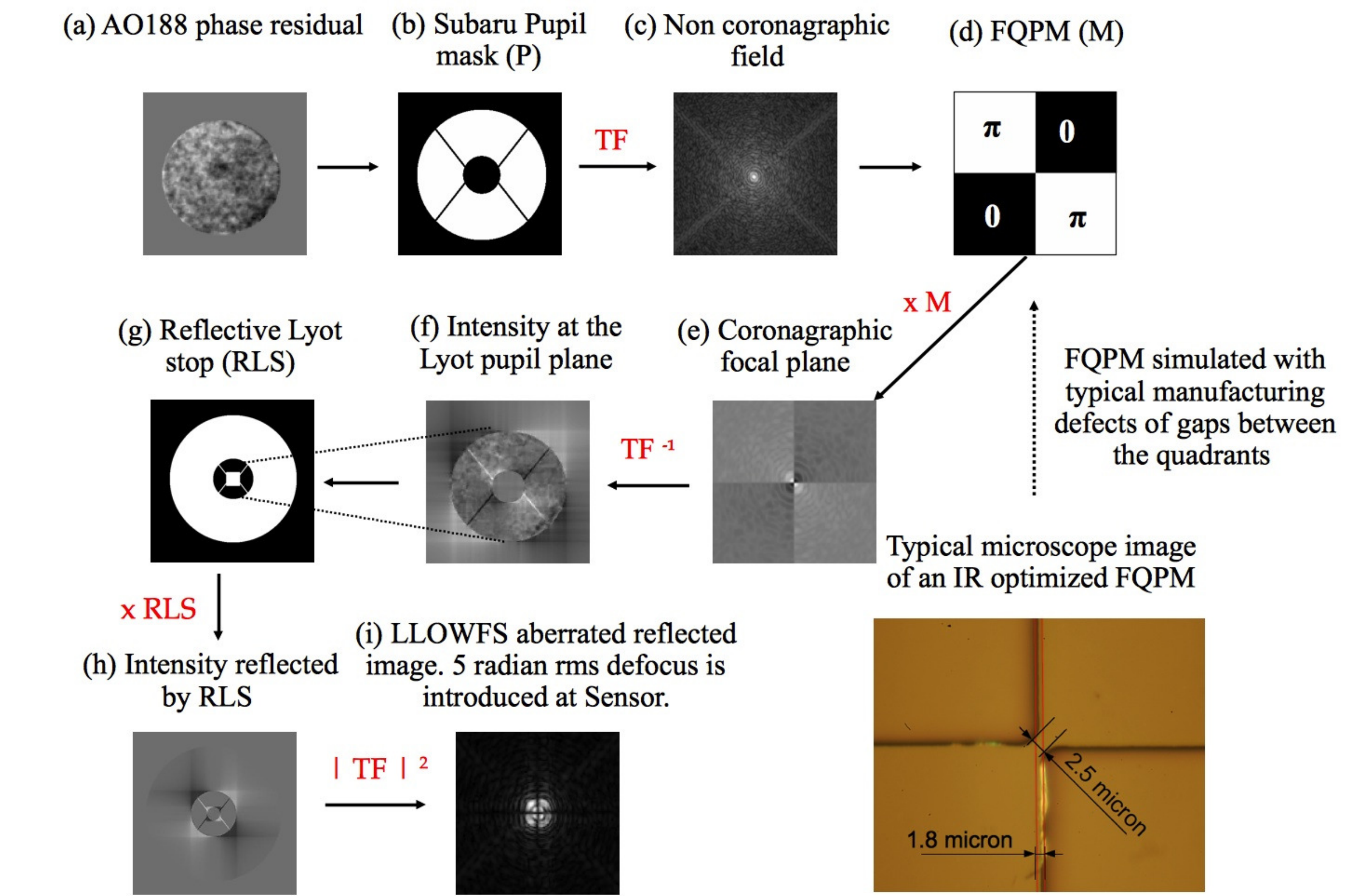}}}
   \caption{
Intensity pattern at each plane of the simplified optical layout (Fig.~\ref {f:plotone}) of the revised SCExAO system at Subaru Telescope showing the Lyot-based low order wavefront sensor (LLOWFS) with the four quadrant phase mask (FQPM) as an example. The input wavefront is the Subaru Telescope's Adaptive Optics System AO188 corrected residual phase map of 150~nm rms at 1.6~$\micron$. The white surface in the reflected Lyot stop (RLS $(g)$) is reflective while the black surface is transmissive. See section \ref{s:OC} for more details.}
  \label{f:plottwo}
\end{figure*}
To make our simulations and experiments relevant to the SCExAO testbed, we used the pupil geometry (central obstruction and spider arms) of the Subaru Telescope. In addition SCExAO is fed by a facility AO system known as AO188, which delivers post correction RMS wavefront errors of $\approx$~200~nm in good seeing (Strehl ratio of $\approx$~40~\% in the H-band). These metrics were included in our simulations and will be used throughout this body of work.
%
\subsection{Pointing error estimation based on linearity approximation}
\label{s:LA}
In a post-AO correction scenario, residual phase errors can be assumed to be small ($\ll$~1 radian of rms wavefront error). 

Simulations, as well as experimental results for the CLOWFS system have demonstrated that for small pointing errors, the intensity fluctuations in the reflective focal plane image are a linear function of low-order phase errors before the coronagraph. Applying the same principle to the LLOWFS, our estimates of the pointing errors are based on the linear relationship between sensor intensity and phase errors. 

If we let $I_{0}$ represent a reflected reference image, acquired by the low-order sensor with no aberrations, and $I_{R}$ the image with some aberrations, we assume that we can relate the difference between these two images by decomposing it into a linear combination of modes. If one considers tip-tilt alone, then we can write:
\begin{equation}
\noindent
I _{R(\alpha_{x},\alpha_{y})} - I_{0} = \alpha_{x} S_{x} + \alpha_{y} S_{y}
\label{eq:1}
\end{equation}
where $S_{x}$ and $S_{y}$ represent the sensor's response to tip and tilt respectively. For any instant image $I_{R}$, one can therefore identify an unknown tip-tilt $(\alpha_x,\alpha_y)$, by direct projection on the basis of modes, or using a least squares algorithm. 
%
%
\section{Lyot-based low order wavefront sensor: Optical elements and realistic simulations}
\label{s:param}
In this section, we describe a simulation tool we have developed to test the functionality of our concept on a low IWA FQPM coronagraph. We discuss the factors that defines the performance criteria of LLOWFS for example cross-talk between modes, linearity range of sensor response and sensitivity towards aberrations upstream of the coronagraph. 
 \begin{figure*}
   \centerline{
        \resizebox{0.7\textwidth}{!}{\includegraphics{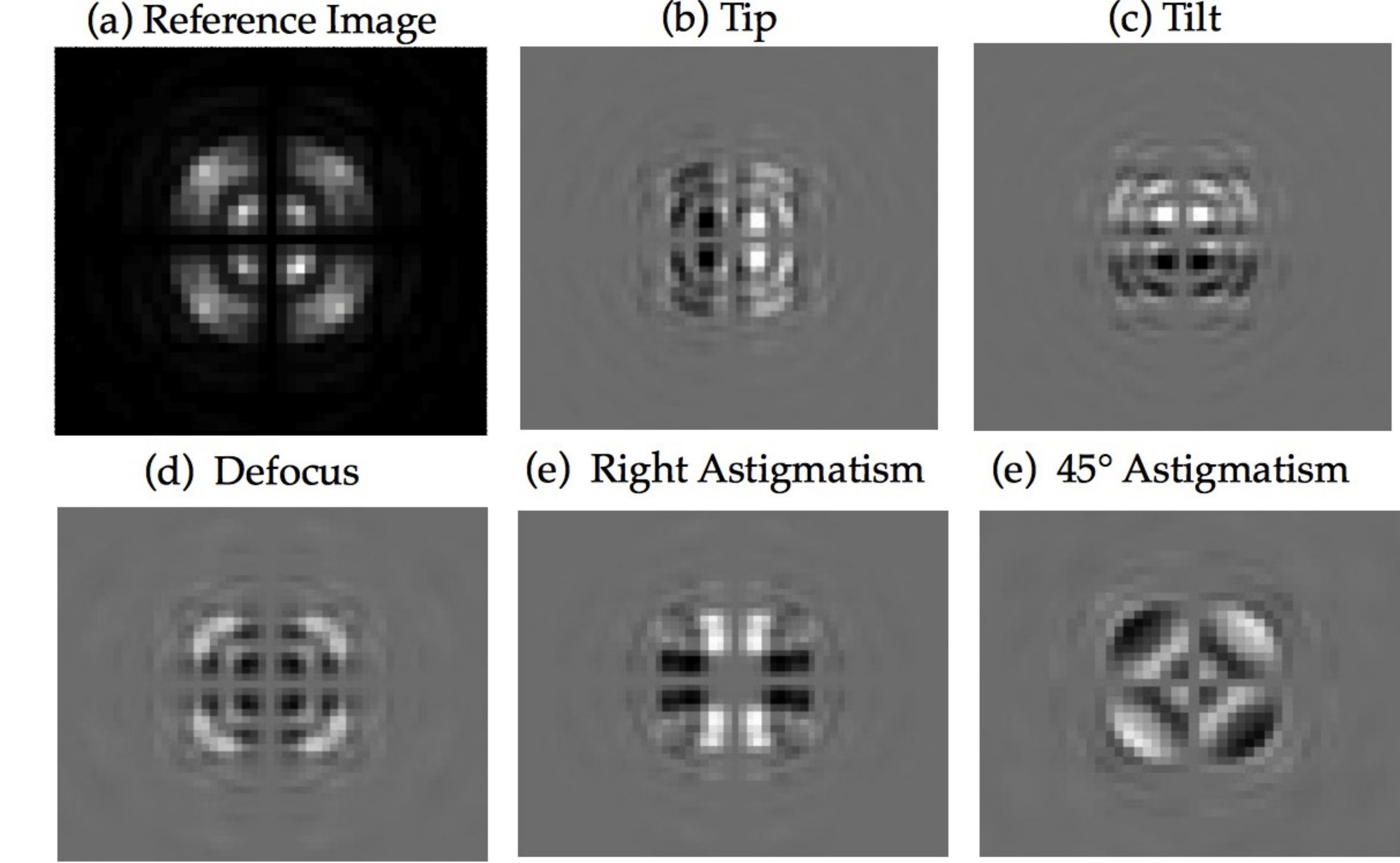}}}
   \caption{
 Simulated response of the LLOWFS sensor to the low-order modes which are obtained after applying 0.01 radian rms phase error per mode in an aberration free system. These images are our calibration frames that we have used to measure the pointing errors present in the aberrated wavefront at the entrance pupil as discussed in detail in section \ref {s:ns}. 5 radian rms of defocus is introduced at the low-order sensor position. The images shown here have the same brightness scale.)}
  \label{f:plotthree}
\end{figure*}
\subsection{Simulation Elements}
\label{s:OC}
A simplified optical layout of LLOWFS is shown in Fig.~\ref{f:plotone} and the corresponding intensity distribution at each plane of a simulation element is shown is Fig.~\ref{f:plottwo}. Below we briefly describe the simulation parameters considered. 

The AO188 corrected beam is focused on a FQPM optimized for 1.6~$\micron$. As shown in Fig. \ref{f:plottwo}(d), the FQPM divides the focal plane in four quadrants and provides a $\pi$-phase shift between adjacent quadrants, resulting in self-destructive interference in the re-imaged pupil. To account for typical manufacturing irregularities, we simulated the FQPM such that it included gaps of 1.8~$\micron$ (0.04~$\lambda/D$) between neighboring quadrants and gaps of 2.5~$\micron$ (0.06~$\lambda/D$) across the diagonal in the center of the mask seen in Fig.~\ref{f:plottwo}(d). These values corresponds to the size of the defects in FQPM used in laboratory experiments. The mask creates a square diffraction pattern around the edge of the central obstruction in the pupil plane as can be seen in Fig.~\ref{f:plottwo}(f). 

The unused diffracted starlight is then reflected via a RLS as shown in Fig.~\ref{f:plottwo}(g). The inner shape of the RLS is a square in order to block the square diffraction pattern around the obstruction. The outer diameter of the reflective area is chosen to be three times the diameter of the pupil in order to collect the majority of the light around the pupil. The image~$(h)$ shows what is seen by the sensor at the RLS plane. 

The image~$(i)$ represents the RLS image ($I_{R}$) which has tip-tilt and other high order modes to be analyzed by the sensor. Note that the transition lines visible in the sensor image are related to the geometry of the FQPM. 5 radians rms of defocus were introduced in the sensor position to ensure that it is larger than the focus term to be measured to avoid sign ambiguities. Other higher order modes can also be measured, however we estimated tip-tilt errors only in this paper.  
%
\subsection{Calibration Procedure}
\label{s:cal}
The LLOWFS is not an absolute pointing system. It is a differential sensor and requires calibration prior to estimating the best centering of the PSF on the coronagraph. To study the behavior of LLOWFS to the unknown tip-tilt errors upstream of the coronagraph, we calibrate the response of the sensor to the aberrations present in the wavefront at the entrance pupil. 

The calibration procedure includes applying a phase map with a controlled amount of tip-tilt and other modes to our system in ideal conditions of no aberration and recording the calibration frames prior of studying the LLOWFS under aberrated conditions.

The sensor response $S_{x}$ and $S_{y}$ towards tip and tilt errors in equation~\ref {eq:1} are estimated by obtaining the respective calibration frames after applying $\approx$~0.01 radian rms of tip and tilt to our system. An example of the response is shown in Fig.~\ref {f:plotthree}, for tip-tilt as well as other modes: defocus, astigmatism etc showing that the LLOWFS is a versatile wavefront sensor which is capable of distinguishing tip-tilt errors from the other aberrations. In all of our simulations, we will use these calibration frames to estimate the tip-tilt errors in the wavefront. 

LLOWFS relies on the shape of the reflected image, not its position on the sensor. However, during the calibration frames acquisition process, if environmental factors (temperature variations, flexure of the instrument) induces tip-tilt, then it will require the recalibration of the system with a new reference image.
%
%
 \begin{figure}[ht]
   \centerline{
        \resizebox{0.5\textwidth}{!}{\includegraphics{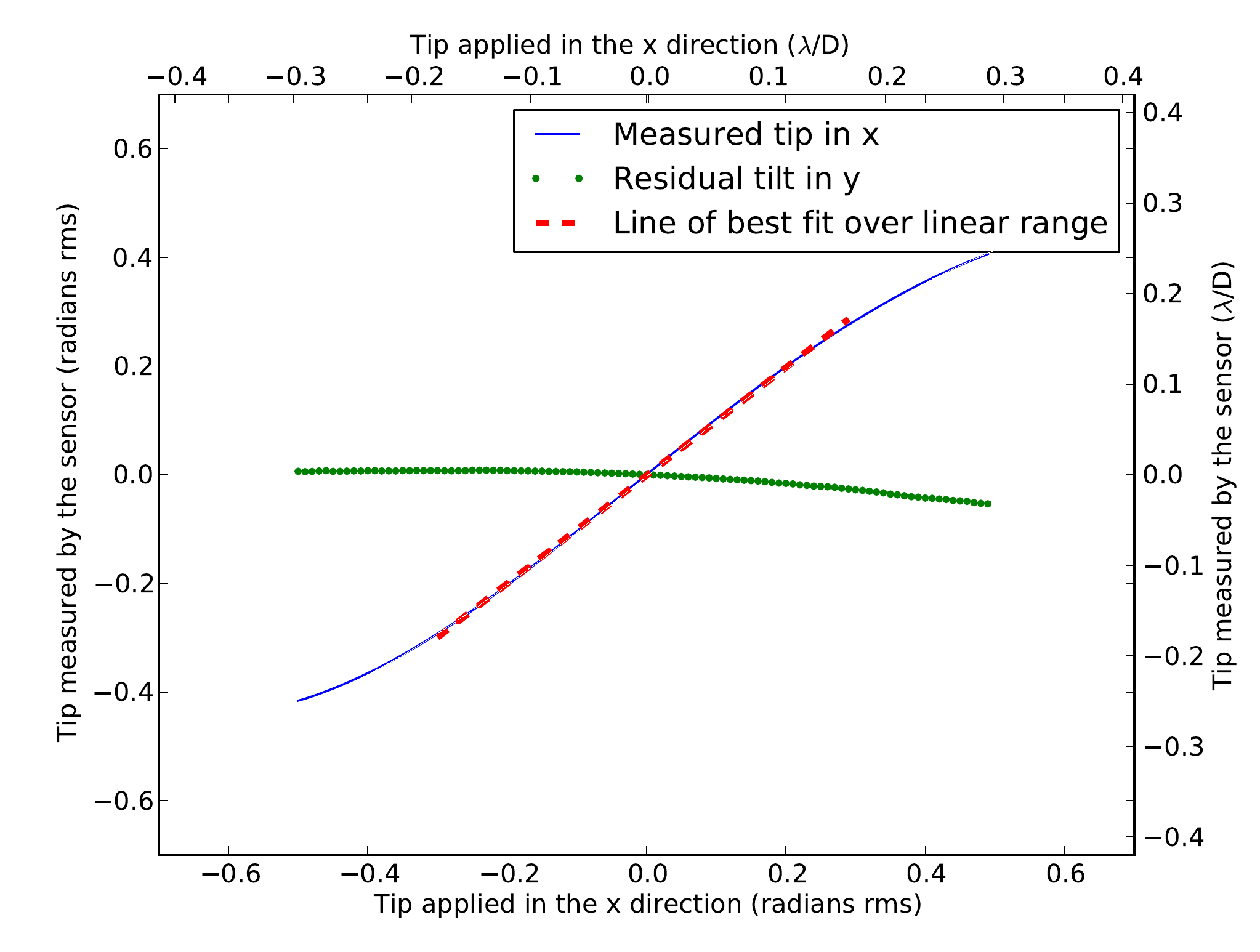}}}
   \caption{Response of the sensor to the tip aberration applied in the x direction at the entrance pupil. The sensor estimates the tip error linearly within $\pm$~0.3~radian rms ($\pm$ 0.18 $\lambda/D$) of phase error with 1\% non-linearity at 0.3 radian rms. The residual tilt in the y direction shows no cross talk between the tip-tilt error measurements. The red dash line is the linear regression done within linearity range i.e. $\pm$~0.3 radians rms.}
  \label{f:plotfour}
\end{figure}
\subsection{LLOWFS Numerical Simulations}
\label{s:ns}
In our simulations, we have studied the LLOWFS without considering photon noise in this paper. 
\subsubsection{Ideal case: aberration-free system}
\label{s:noph}
In the basic LLOWFS configuration shown in Fig.~\ref{f:plotone}, we considered no phase defects at the entrance pupil. First we acquired the on-axis image of the reflected light and recorded it as our reference image $I_{0}$ as shown in Fig.~\ref{f:plotthree}(a). 

Next we applied a varying amount of tip between $\pm$~0.5~radians rms ($\pm$~0.3~$\lambda/D$) in the x direction with a step size of 0.01 radian rms and recorded a cube of 100 images as $I_{R\alpha_{x}}$.  We then estimated tip-tilt aberrations $\alpha_{x}$ and $\alpha_{y}$ present in $I_{R\alpha_{x}}$using equation~\ref {eq:1} by solving them as a least squares problem. 

The applied tip with respect to the estimated tip is shown in Fig.~\ref {f:plotfour}. The residual tilt in the y direction which is almost negligible is also shown. The sensor showed a linear response to tip/tilt aberrations within $\pm$~0.3~radians rms ($\pm$ 0.18 $\lambda/D$) of phase error with 1~\% non-linearity in measurement at 0.3~radian rms.  

One of the properties of the LLOWFS technique is that it measures the modes independently from each other. There is no cross talk between measured modes in the low order aberration regime and the reconstruction of the image should work if the estimation of errors are done in the linear regime, more details of which can be found in CLOWFS paper \citet{Guyon2009}.
\subsubsection{Sensor linearity under low-order phase errors}
\label{s:phdef}
In this section, we study the response of the sensor to tip-tilt errors under the influence of multiple low-order phase aberrations such as tip, tilt, focus and astigmatism simultaneously. We consider an input wavefront as a cube of 100 phasemaps (image cube $I_{R}$) of low order modes. Each phasemap has 5 low order errors with rms amplitude values increasing sequentially between $\pm$~0.5~radians. For example the first phasemap has all the modes with -0.5 radians (290~nm total phase error over the Subaru pupil), the $50^{th}$ phasemap is free of aberrations and the $100^{th}$ phasemap has + 0.5 radians amplitude per mode.

\begin{figure}[ht]
   \centerline{
        \resizebox{0.5\textwidth}{!}{\includegraphics{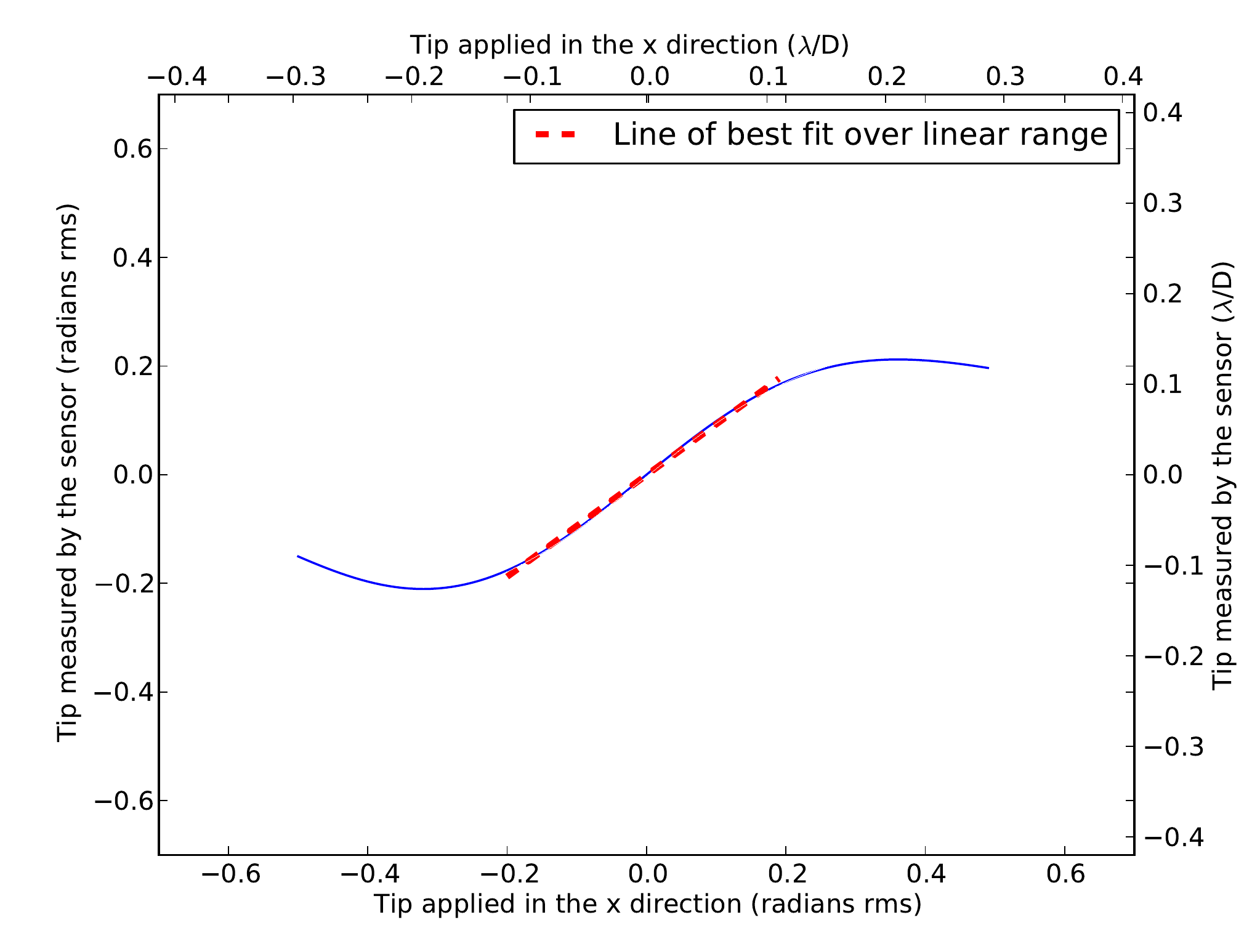}}}
   \caption{Linearity check of the sensor response to tip errors present in low-order phasemaps applied at the entrance pupil. The x-axis displays the tip aberrations applied in each phase map of the image cube. The corresponding y axis shows the tip errors estimated by the LLOWFS. Under the influence of multiple low order aberrations, the sensor response becomes non-linear beyond\\ $\pm$~0.2 radians rms ($\pm$ 0.12 $\lambda/D$).}
  \label{f:plotfive}
\end{figure}

Using equation~\ref {eq:1}, we estimated the tip-tilt coefficients through least squares methods and we present only the measured tip in the x direction in Fig.~\ref {f:plotfive}. The plot showed compares the amount of tip present in the image cube with respect to the tip measured by the sensor. In the presence of low-order aberrations, the sensor response is linear for tip aberrations between $\pm$~0.2 radians rms ($\pm$~0.12~$\lambda/D$) with 14\% non-linearity in measurement at 0.2 radian rms and 32\% non-linearity at 0.3 radian rms. Unlike the single aberration case, the system response beyond $\pm$~0.2 radians rms range becomes non-linear in presence of multiple low order aberrations and therefore makes the loop diverge. 

We have repeated the tip-tilt error measurement analysis for different defocus positions in the sensor and noticed that the sensor linearity remains unaffected. 
%
%
%
%
\subsubsection{Post AO188 phase residuals as an input wavefront to the LLOWFS}
\label{s:sim1}
The next step is to check the sensor linearity range and sensitivity towards high-order aberrations present in the residuals of the Subaru Telescope facility AO system. This is done by simulating a series of 200 post AO188 phasemaps with unknown low and high order aberrations with $\approx$~180~nm rms amplitude. Fig.~\ref{f:plottwo}(a) represents the first phase map of the series. 

 \begin{figure}[ht]
   \centerline{
        \resizebox{0.5\textwidth}{!}{\includegraphics{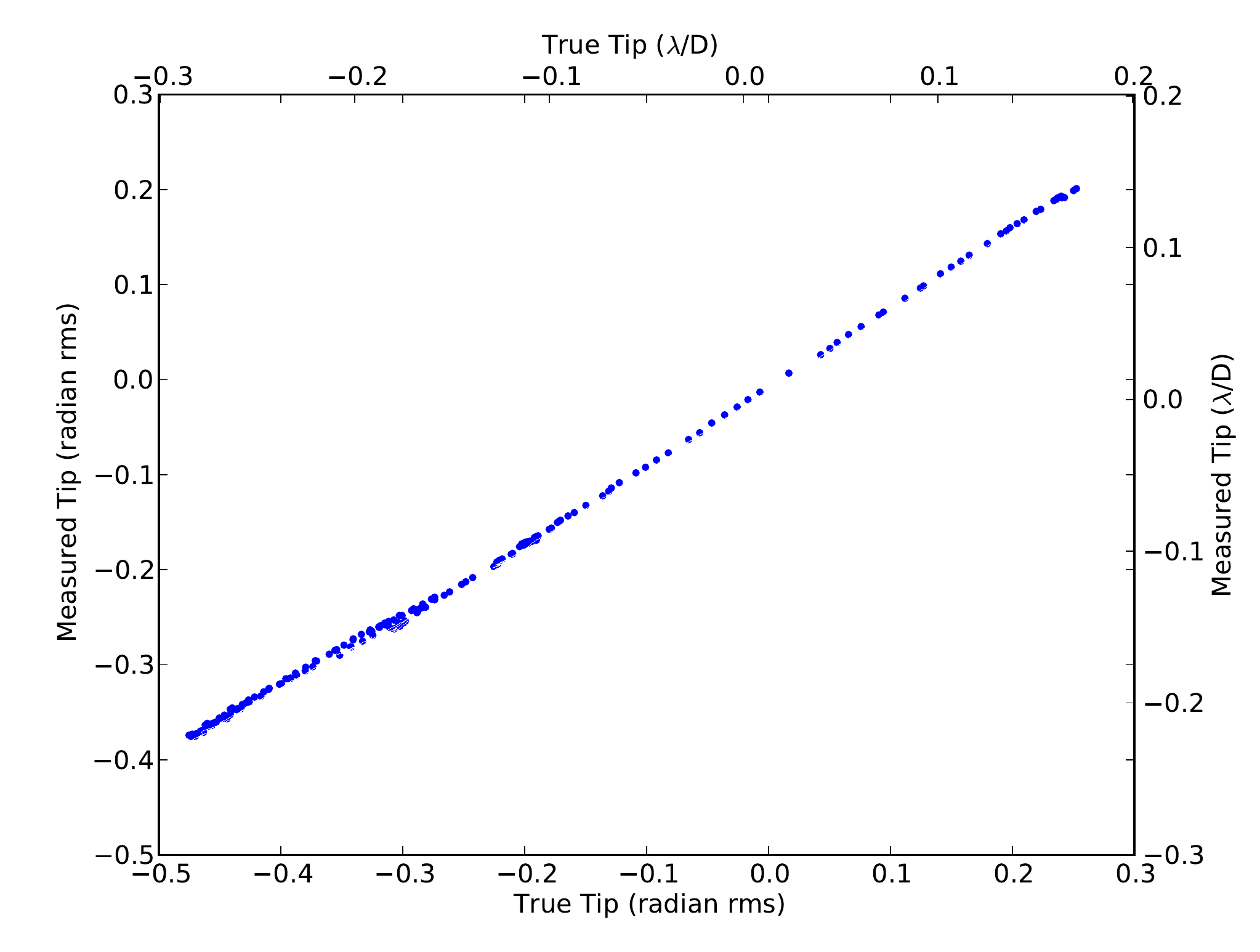}}}
   \caption{Linearity of sensor response to tip aberrations present in AO188 phase residuals. The x-axis represents phase error contributed by tip aberrations in each phasemap. The corresponding y-axis shows the response of the sensor to these tip errors.}
  \label{f:plotsix}
\end{figure}
\begin{figure*}
   \centerline{
        \resizebox{1.0\textwidth}{!}{\includegraphics{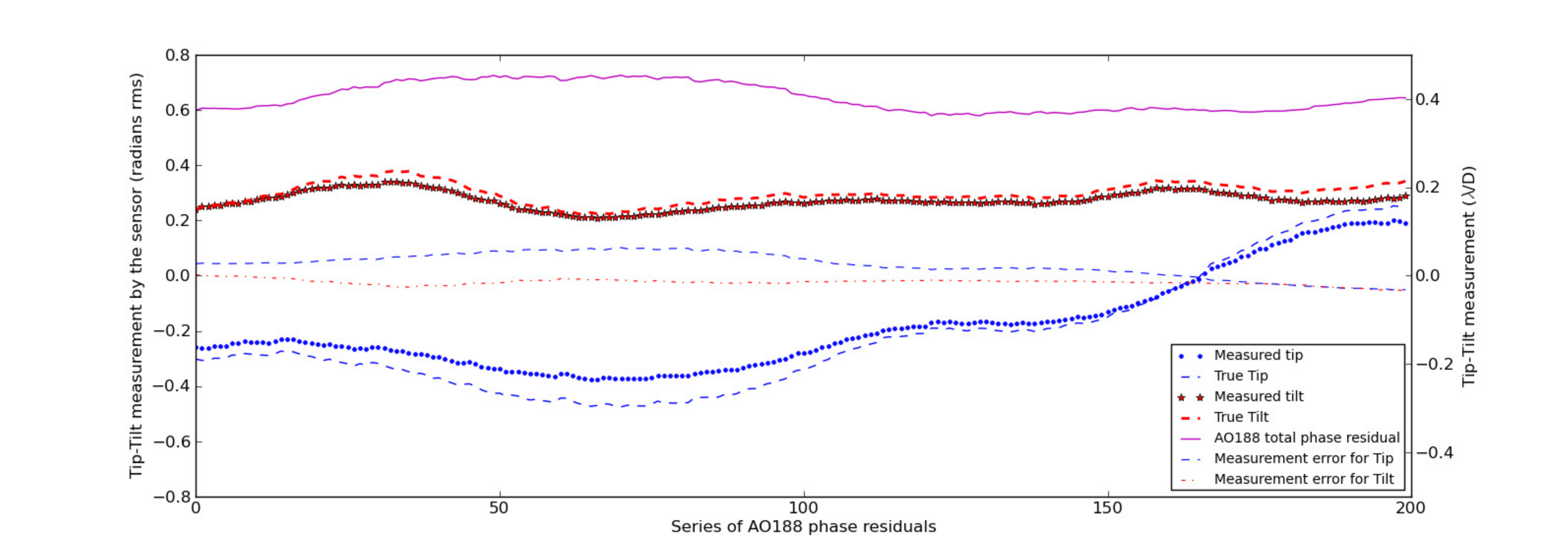}}}
   \caption{
(i) Simulated sensor response to tip-tilt errors present in the realistic phase residuals of the Subaru Telescope AO188. The simulated series is comprised of 200 post AO phasemaps with unknown high and low order modes with $\approx$~180~nm phase rms over the Subaru pupil. In the presence of high order modes and large tip-tilt errors, the predicted linearity of sensor decreases in the open loop regime. The measurement errors for tip and tilt shown are the difference between the measured and true tip/tilt errors. The standard deviation of this difference shows sensor's measurement accuracy of $\approx~10^{-2}~\lambda/D$ per mode.}
  \label{f:plotseven}
\end{figure*}
\begin{figure*}
   \centerline{
        \resizebox{0.7\textwidth}{!}{\includegraphics{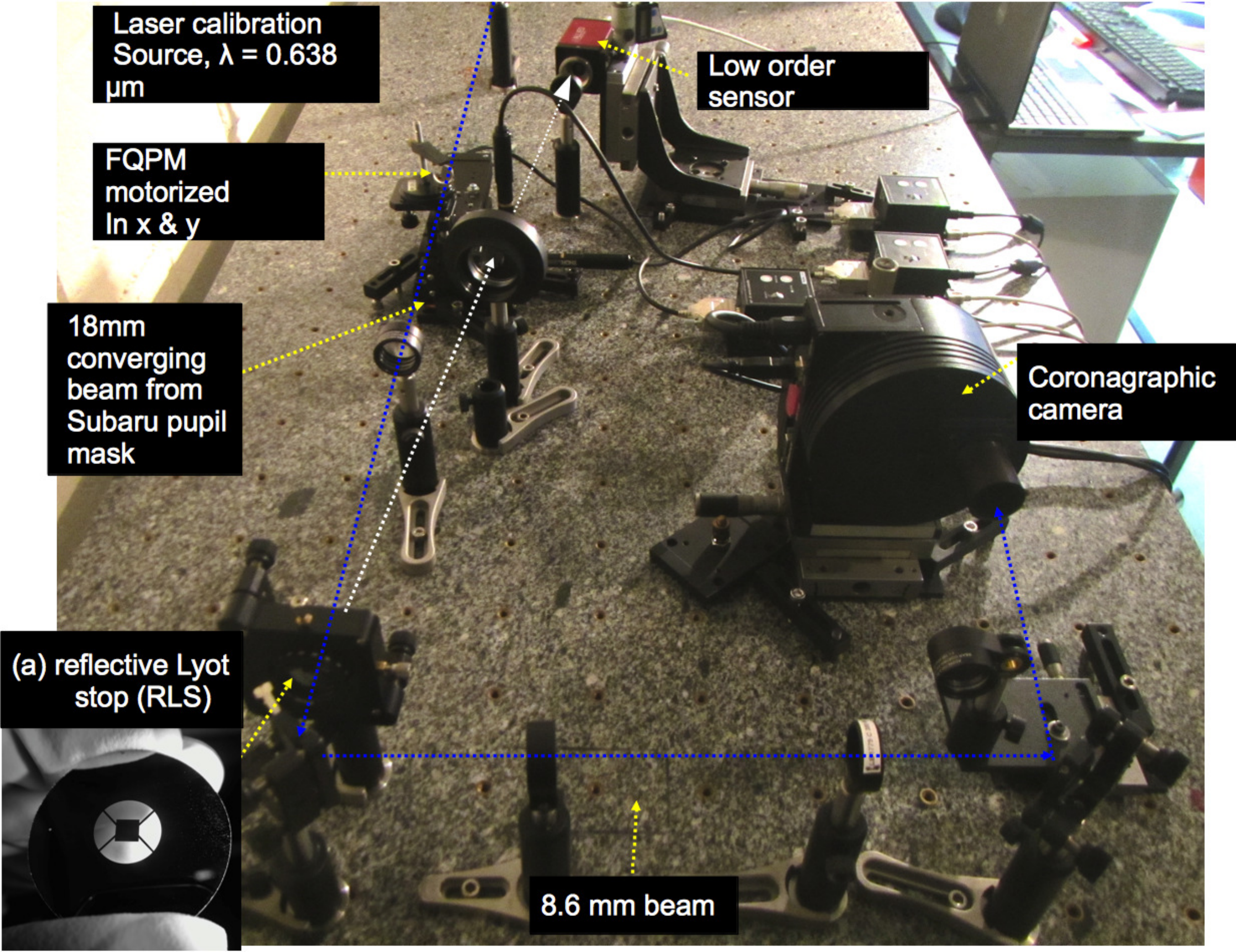}}}
   \caption{Laboratory setup of LLOWFS at Laboratoire d'\'etudes Spatiales et Instrumentation en Astrophysique (LESIA), Observatoire de Paris-Meudon (June 2013).}
  \label{f:ploteight}
\end{figure*}
With the AO188 residual phasemap series as an input wavefront to our system, we measured the tip and tilt errors. Figure~\ref {f:plotsix} plots the true tip aberrations present in the residual phasemaps versus tip aberrations measured by LLOWFS low-order sensor. 

We also compare the sensor response to tip-tilt errors with their actual residual value for 200 phasemaps as shown in Fig.~\ref {f:plotseven}. We notice that for small tip-tilt excursions, the sensor response to low-order aberrations provides a reliable measurement of the tip-tilt. The fidelity of the reconstruction degrades for larger tip-tilt excursion (beyond $\pm$~0.2 radian rms), due to non-linearity effects, but the sensor remains well behaved, and in closed loop, would converge toward the reference position, despite the non-linearity. Under realistic simulations, the sensor delivered a measurement accuracy of $\approx~10^{-2}~\lambda/D$ (2-12 nm at 1.6~$\micron$)  per mode as shown in Fig.~\ref {f:plotseven}. 

The higher order aberrations and large tip-tilt errors can affect the LLOWFS performance by introducing crosstalk between modes. Astigmatisms can be misinterpret with higher order mode such as trefoil. We aim to study the behavior of sensor response to low-order aberrations under the influence of higher order aberrations in future work.
%
\section{LLOWFS Laboratory Implementation}
\label{s:lab}
 An early version of the LLOWFS has been implemented on the coronagraphic testbed at LESIA, Observatoire de Paris. The optical design of our experiment is shown in Fig.~\ref{f:plotone} and the laboratory setup is represented in Fig.~\ref {f:ploteight} which included:
\begin{itemize}
\item{Laser source of $\lambda$~=~0.638~$\micron$}
\item{Subaru pupil mask}
\item{FQPM (optimized at $\lambda$~=~0.635~$\micron$), mounted on two motors allowing it to move in the x and y directions.}
\item{RLS (8.6 mm Lyot outer pupil diameter with 25.4~mm of reflective annulus around it) placed at an angle of 8$\degree$ to reflect the light towards the low order sensor. 

The RLS is a fused silica disk of 1.5 mm thickness as shown in Fig.~\ref{f:ploteight}(a). The substrate flatness is better than 5~$\micron$. The black region is reflective chrome with 60~\% reflectivity in near IR (1200 nm) while the white region is transparent.}  
\item{Low-order sensor (pixel size = 7~$\micron$). We introduced defocus in the sensor position and we estimated the value to be 5~$\pm$ 0.3 radians rms. } 
\end{itemize}
\begin{figure}
   \centerline{
        \resizebox{0.5\textwidth}{!}{\includegraphics{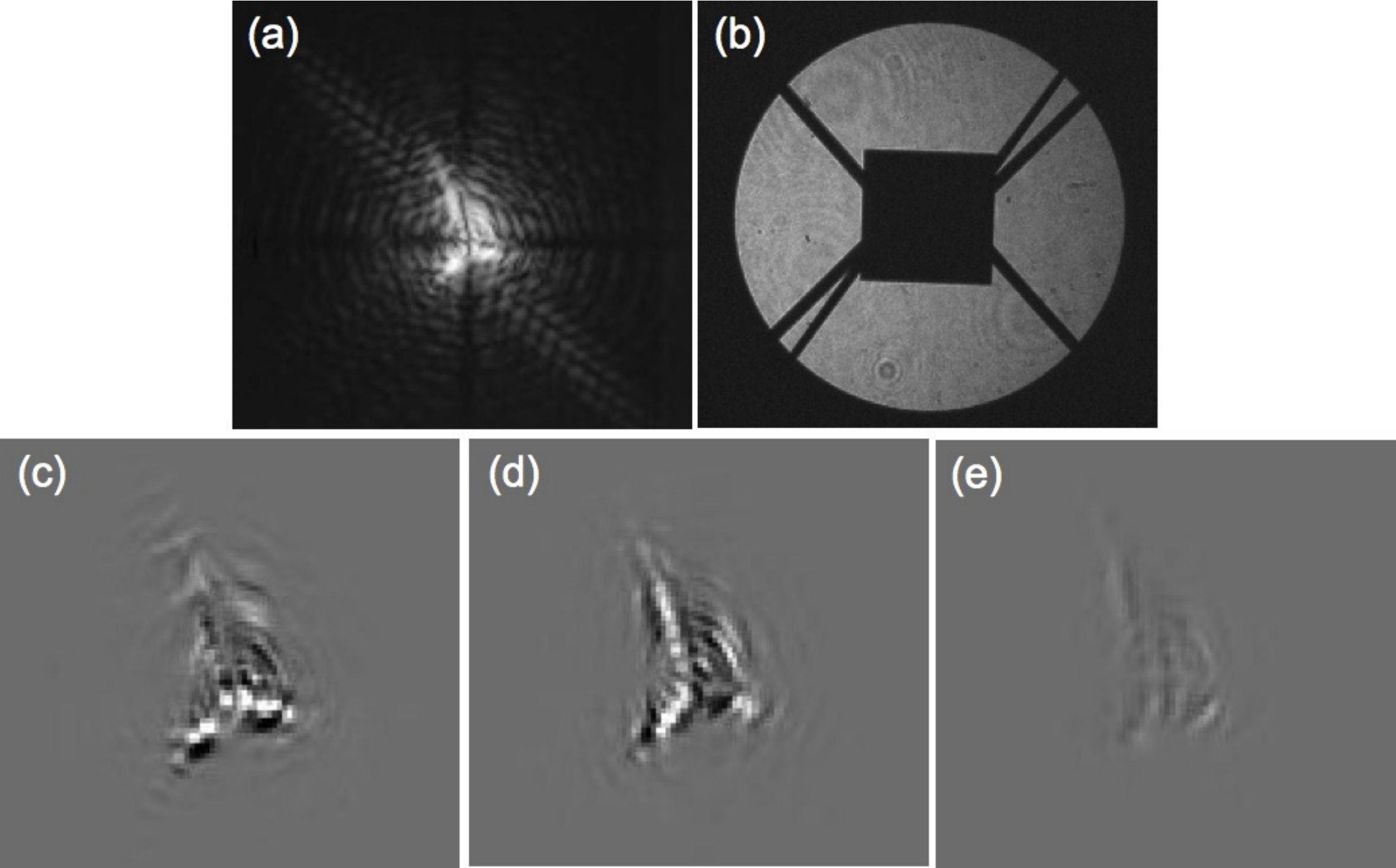}}}
   \caption{(a) Reference image acquired by LLOWFS. (b) RLS overlapped with the Subaru pupil mask. (c) Calibration frame for tip (0.12~$\lambda/D$). (d) Calibration frame for tilt (0.12~$\lambda/D$). (e) Difference between two reference images acquired before and after recording the mode measurements. (Exposure time for recording the images : 500~ms. The images shown here have the same brightness scale.)}
  \label{f:plotnine}
\end{figure}
 \begin{figure}[ht]
   \centerline{
        \resizebox{0.5\textwidth}{!}{\includegraphics{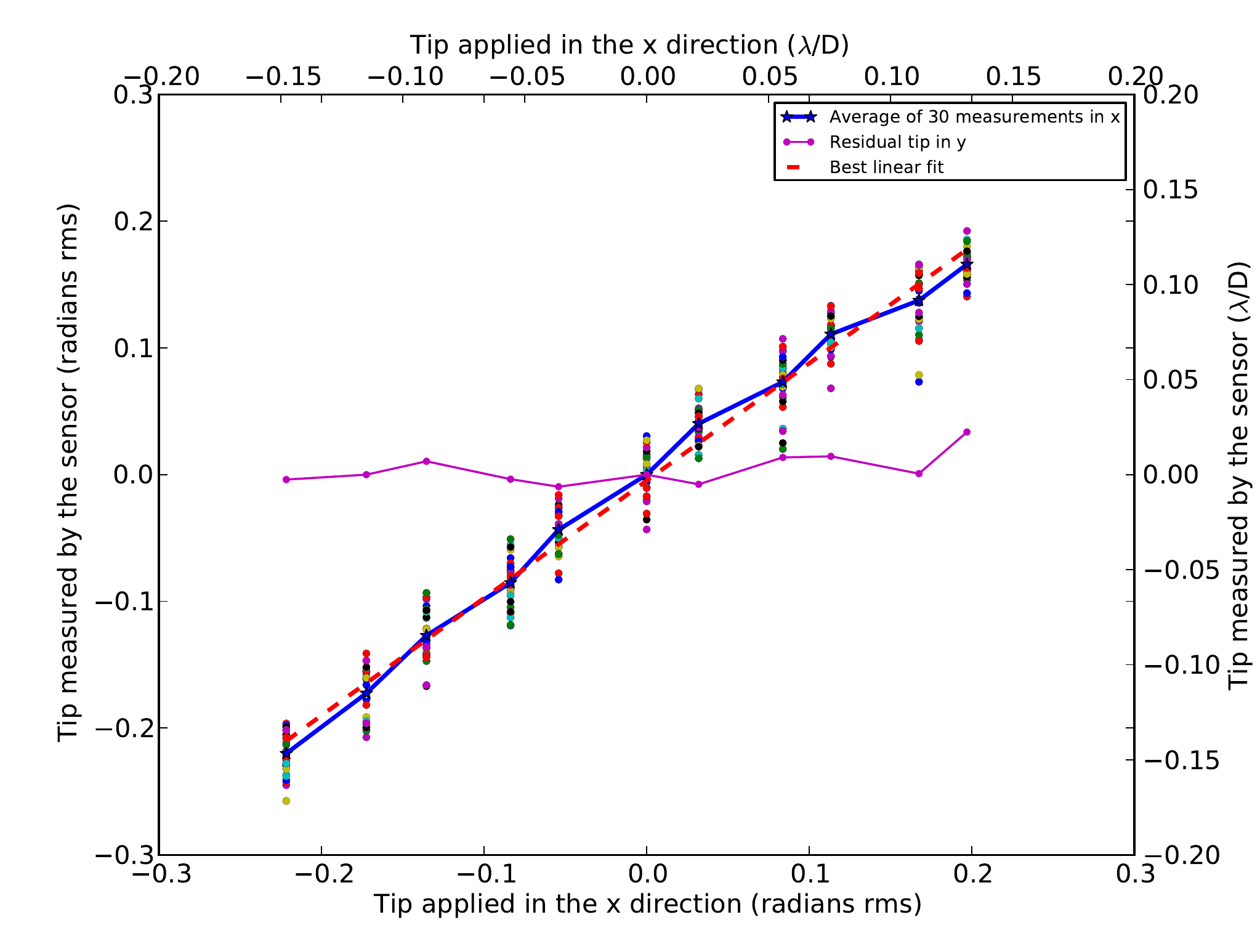}}}
   \caption{Laboratory results for tip errors : Graph shows the linearity of sensor response to the tip aberration applied to the FQPM in the x direction. The residual tilt in the y direction stays around zero for all the measurements for tip errors showing no cross talk between the modes. For each step, 30 images are taken showing dispersion of the measurement due to local turbulence. The red dash line is the linear regression done in the linearity range i.e. within $\pm$~0.12~$\lambda/D$ ($\pm$~0.19 radians rms). The sensor measurement accuracy for tip errors is 0.009~$\lambda/D$.}
  \label{f:plotten}
\end{figure}
 \begin{figure}[ht]
   \centerline{
        \resizebox{0.5\textwidth}{!}{\includegraphics{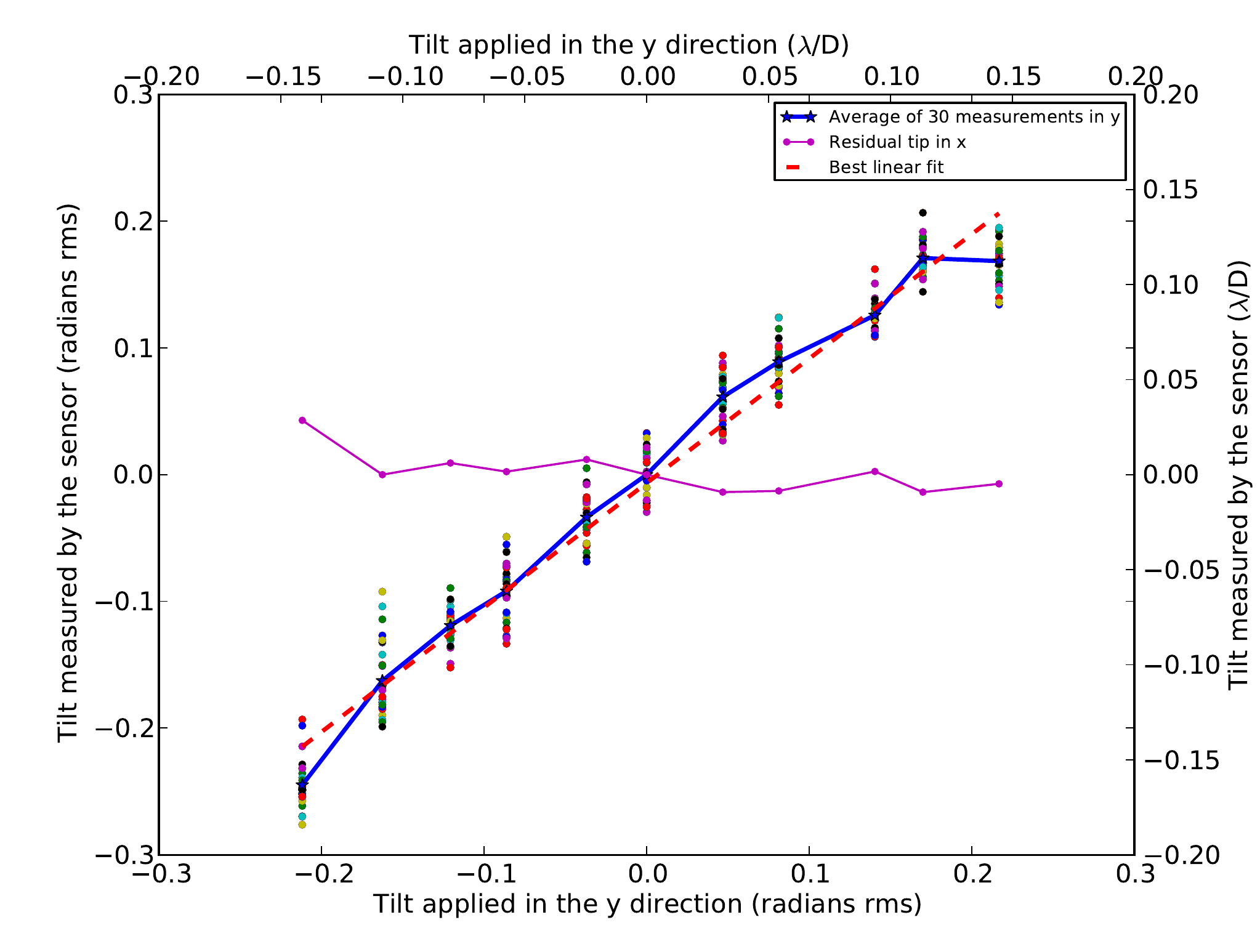}}}
   \caption{Laboratory results for tilt errors: Graph shows the linearity of sensor response to the tilt aberration applied in the y direction. The residual tip in x measured by the sensor stays around zero. The red dash line is the linear regression done in the linearity range i.e. within $\pm$~0.12~$\lambda/D$ ($\pm$~0.19 radians rms). The sensor measurement accuracy for tilt errors in the y-direction is 0.011~$\lambda/D$.}
  \label{f:ploteleven}
\end{figure}
%
\subsection{Procedure of acquiring the measurements}

First the FQPM is aligned and the rejection of $\approx$~30 has been obtained. The reasons for such sub-optimal performance are: wavefront distortions, manufacturing accuracy of the mask, the FQPM rejection wavelength was not optimized for the laser source, in addition, the spiders and the secondary of the pupil affected the extinction capability of the coronagraph.

The intensity reflected by RLS is recorded as a reference image $I_{0}$ as shown in Fig.~\ref{f:plotnine}(a). Image~\ref{f:plotnine}(b) shows the RLS having manufacturing defects: the angle between adjacent spider arms were 10$\degree$ less than required. This meant that we could not mask all of the spider arms of the pupil, the traces of which is clearly visible in the reference image.

The FQPM is then moved in x with a step of 0.12~$\lambda/D$ and the response of the sensor is acquired for tip as shown in Fig.~\ref{f:plotnine}(c). The sensor response obtained for the FQPM for the same step in the y direction is shown in Fig.~\ref {f:plotnine}(d). 

Note that our approach of obtaining sensor measurements in the x and y direction were identical. We registered the calibration frame for tip and tilt separately prior to testing the linearity of the sensor under large excursions in tip-tilt errors.

We then moved the FQPM in x within $\pm$~0.2~$\lambda/D$ ($\pm$~0.32~radians rms) with a step size of $\approx$~0.05~$\lambda/D$ and stored the image cube as $I_{Rx}$. We recorded two reference frames before and after taking our measurement. In Fig.~\ref{f:plotnine}(e), the difference clearly shows structure that can be explained by the local turbulence in the laboratory and thermal drifts of the mechanical elements driving the FQPM. 

Figure~\ref{f:plotten} and \ref{f:ploteleven} shows the response of the sensor to tip and tilt in x and y studied independently. For every step applied in the x or y, a series of 30 images were recorded by the sensor. The dispersion in the data is a result of the local turbulence. For tip aberrations applied in the x direction, the tilt aberrations in the y direction stayed around zero. So we confirm the cross-talk independency between tip-tilt error measurements as predicted in the simulation.

As can be seen, the sensor has a linear response over the majority of the aberration amplitude range tested with some possible non-linearity creeping in at the beginning and end of the range under test. However, without more data this can not be concluded with any certainty. Nonetheless this means the sensor can be used over a range of $\pm$~0.12~$\lambda/D$ ($\approx$~0.19 radians phase rms over Subaru pupil) reliably.
%
\subsection{Problems faced during LLOWFS laboratory implementation}
\label{s:disc}
The implementation of the LLOWFS in the laboratory was limited by various factors such as:
\begin{itemize}
\item{Local Turbulence}
\item{Non-optimized alignment of the testbed}
\item{Drift in the reference image of the LLOWFS over time}
\item{The substrate quality of RLS and its manufacturing defects}
\item{AR coatings on the FQPM were not optimized for the phase mask working wavelength }
\item{Precision of the encoders on the stages driving the FQPM ($10^{-2}~\lambda/D$)}
\item{Low-order sensor read out noise (15e-)}
\end{itemize}

Even with all the limiting factors listed above, the LLOWFS implementation in the laboratory has efficiently demonstrated in open loop measurements an accuracy of $\approx~10^{-2}~\lambda/D$  per mode for the FQPM at 638 nm. 

\section{General Discussion}
\label{s:con}
The Lyot-based low order wavefront sensor (LLOWFS) is highly photon efficient, as for pointing control it uses the starlight diffracted outside the phase mask which is otherwise wasted in a conventional phase mask coronagraph (PMC). Our technique essentially requires a reflective Lyot stop and a low-order sensor which is simply a detector. The LLOWFS schematic layout presented in Fig.~\ref {f:plotone} is adaptable to any PMC.

We showed that the LLOWFS measurement is not only limited to tip-tilt as we could also potentially sense defocus and astigmatism which are distinguishable from tip-tilt signals as shown in Fig.~\ref{f:plotthree}. However the shortcomings of the LLOWFS that we noticed are: the maximum amplitude of tip/tilt excursions that can be measured is limited, there is a high risk of misinterpretation of astigmatism with higher order modes such as trefoil, the sensor performance is influenced under high order aberrations which can introduce cross talk between modes. One more aspect that we have not studied yet is the photon sensitivity versus the sensor defocus position. The amount of defocus that we introduce will likely affect the sensitivity of the LLOWFS. We therefore aim to study the operation of the LLOWFS under the circumstances listed above in future work. 

Our system can also be adapted as a post-processing technique \citep{Vogt2011} to enhance the sensitivity of the coronagraph. The image reconstructed through the LLOWFS can be subtracted from the science image to calibrate out the low-order residuals. 

The LLOWFS concept is reliable for high performance small inner working angle coronagraphs which provides $10^{-2}~\lambda$/D of pointing measurement accuracy, hence efficiently preventing the coronagraphic leaks which are the main cause of the degradation in the nulling performance. The combination of small IWA PMC + sub mili-arcsecond level pointing stability of the LLOWFS applied to extremely large telescopes could enable the direct imaging of reflected light habitable zone planets. 

The LLOWFS is indeed a necessity to control the pointing errors with PMC and an appealing solution for not only ground-based but for space-based telescopes as well.

\section{Conclusion}
\label{s:fd}
High throughput, low inner working angle (IWA) coronagraphs are essential to directly image and characterize (spectroscopy) exoplanets. However their performance is affected due to lack of accurate pointing control. We addressed this issue by introducing a robust, easily adaptable technique to prevent the coronagraphic leaks for low IWA phase mask coronagraphs.

In this paper, we showed in simulations that with the realistic AO phase residuals as an input, the Lyot-based low order wavefront sensor (LLOWFS) by using the unused starlight reflected by the Lyot stop is capable of measuring tip, tilt with the accuracy of $\approx~10^{-2}~\lambda/D$ (2-12~nm at 1.6~$\micron$) per mode on the four quadrant phase mask (FQPM). 

We demonstrated the LLOWFS preliminary laboratory implementation and performance under a lack of stable environment. We estimated tip-tilt errors with measurement accuracy of $\approx~10^{-2}~\lambda/D$ at 638~nm per mode in an open loop regime with the FQPM. 

Our simulation and laboratory performance shows that the LLOWFS has reliable accuracy with the FQPM. Future work includes the simulation, performance verification and contrast sensitivity of our technique with other phase masks such as the Roddier \& Roddier, the vortex, the eight octant phase mask and phase-induced amplitude apodization with a variable focal plane mask. 

Aiming to make the LLOWFS more efficient and versatile, the measurement and correction of other low order aberrations such as focus and the astigmatisms will also be studied.

To further demonstrate our concept we have equipped the revised SCExAO testbed with the phase mask coronagraphs listed above together with their corresponding reflected Lyot stops. The laboratory tests under closed loop regime are currently ongoing and we aim to test the system on-sky soon. 

\end{document}